\documentclass[conference]{IEEEtran} 
\IEEEoverridecommandlockouts

\usepackage{amsmath,amsfonts}
\usepackage{array}
\usepackage[caption=false,font=footnotesize]{subfig}
\usepackage{stfloats}
\usepackage{url}
\usepackage{verbatim}
\usepackage[acronym]{glossaries}
\usepackage{algorithmic}
\usepackage{amsmath,amssymb,amsfonts}
\usepackage{cite}
\usepackage{enumitem}
\usepackage{listings}
\usepackage[shortcuts]{extdash}
\usepackage{textcomp}
\usepackage[dvipsnames]{xcolor}
\usepackage[shortcuts]{extdash}
\usepackage{datetime}
\newdateformat{usdateformat}{\twodigit{\THEMONTH}/\twodigit{\THEDAY}/\THEYEAR}

\usepackage{balance}

\usepackage{graphicx}
\usepackage{booktabs}
\usepackage[referable]{threeparttablex}
\usepackage{tabularx}
\usepackage{multirow}
\usepackage{colortbl}
\renewlist{tablenotes}{enumerate}{1}
\makeatletter
\setlist[tablenotes]{label=\tnote{(\alph*)},ref={(\alph*)},itemsep=\z@,topsep=\z@skip,partopsep=\z@skip,parsep=\z@,itemindent=\z@,labelindent=\tabcolsep,labelsep=.2em,leftmargin=*,align=left,before={\footnotesize}}
\makeatother

\newlist{rdescription}{description}{1}
\AtBeginEnvironment{rdescription}{%
\renewcommand*\descriptionlabel[2][Des]{\hspace\labelsep\eqmakebox[Des][r]{\hfill\normalfont\bfseries #2}}\setlist[rdescription]{leftmargin=\dimexpr\eqboxwidth{Des}+\labelsep}}%

\usepackage[free-standing-units,per-mode=repeated-symbol,binary-units=true,detect-weight=true,detect-family=true]{siunitx}[=v2]

\usepackage[draft]{hyperref}
\usepackage{orcidlink}
\usepackage[noabbrev, capitalise,nameinlink]{cleveref}
\usepackage[stretch=50,shrink=50]{microtype}
\usepackage[pscoord]{eso-pic}

\usepackage{svg}
\usepackage{tikz}

\DeclareSIUnit\squaredkm{\si{km.s^{2}}}
\DeclareSIUnit\devspersquaredkm{dev/\si{km^{2}}}
\DeclareSIUnit\core{core}
\DeclareSIUnit\request{req}
\DeclareSIUnit\cycle{cycle}
\DeclareSIUnit\bps{bps}
\DeclareSIUnit\Bps{Bps}
\DeclareSIUnit\terabps{T\bps}
\DeclareSIUnit\gigabps{G\bps}
\DeclareSIUnit\megabps{M\bps}
\DeclareSIUnit\op{OP}
\DeclareSIUnit\ops{OPS}
\DeclareSIUnit\flop{FLOP}
\DeclareSIUnit\flops{FLOPS}
\DeclareSIUnit\teraops{TOPS}
\DeclareSIUnit\gate{GE}
\DeclareSIUnit\ge{GE}
\DeclareSIUnit\mhz{MHz}
\DeclareSIUnit\ghz{GHz}
\DeclareSIUnit\ipc{IPC}
\DeclareSIUnit\ips{IPS}
\DeclareSIUnit\bits{bits}
\DeclareSIUnit[number-unit-product = ]\percent{\%}

\definecolor{grey0}{HTML}{1B2B34}
\definecolor{grey1}{HTML}{343D46}
\definecolor{grey2}{HTML}{4F5B66}
\definecolor{grey3}{HTML}{65737E}
\definecolor{grey4}{HTML}{A7ADBA}
\definecolor{grey5}{HTML}{8CABA8}
\definecolor{grey6}{HTML}{BECBD2}
\definecolor{grey7}{HTML}{E9EEF0}
\definecolor{red}{HTML}{A8322D}
\definecolor{orange}{HTML}{D97531}
\definecolor{yellow}{HTML}{F8BD3F}
\definecolor{green}{HTML}{92BA51}
\definecolor{teal}{HTML}{108A7C}
\definecolor{blue}{HTML}{2A4857}
\definecolor{purple}{HTML}{875A78}
\definecolor{brown}{HTML}{805A4D}
\definecolor{color1}{HTML}{f94144}
\definecolor{color2}{HTML}{f9c74f}
\definecolor{color3}{HTML}{90be6d}
\definecolor{color4}{HTML}{43aa8b}
\definecolor{color5}{HTML}{577590}
\definecolor{PulpGreen}{HTML}{168638}
\definecolor{PulpBlue}{HTML}{1269b0}
\definecolor{PulpRed}{HTML}{a8322c}
\definecolor{PulpYellow}{HTML}{f2c100}

\PackageWarning{Citation missing:}{#1!}

\PackageWarning{TODO:}{#1!}

\usepackage[normalem]{ulem}

\newacronym{5G}{5G}{5th Generation}
\newacronym{6G}{6G}{6th Generation}
\newacronym{SDR}{SDR}{Software Defined Radio}
\newacronym{TTI}{TTI}{Transmission Time Interval}
\newacronym{RAN}{RAN}{Radio-Access-Networks}
\newacronym[longplural={Scratchpad Memories}]{SPM}{SPM}{Scratchpad Memory}
\newacronym{ACE}{ACE}{AXI Coherent Extensions}
\newacronym{AI}{AI}{Artificial Intelligence}
\newacronym{AMBA}{AMBA}{Advanced Microcontroller Bus Architecture}
\newacronym{AMX}{AMX}{Advanced Matrix Extension}
\newacronym{APB}{APB}{Advanced Peripheral Bus}
\newacronym{API}{API}{Application Programming Interface}
\newacronym{ASIC}{ASIC}{Application-Specific Integrated Circuit}
\newacronym{AVX}{AVX}{Advanced Vector Extension}
\newacronym{AXI}{AXI}{Advanced eXtensible Interface}
\newacronym{BLAS}{BLAS}{Basic Linear Algebra Subprograms}
\newacronym[longplural={Core Complexes}]{CC}{CC}{Core Complex}
\newacronym{CHI}{CHI}{Coherent Hub Interface}
\newacronym{CMOS}{CMOS}{Complementary Metal-Oxide-Semiconductor}
\newacronym{CNN}{CNN}{Convolutional Neural Network}
\newacronym{CPU}{CPU}{Central Processing Unit}
\newacronym{CSR}{CSR}{Control and State Register}
\newacronym{CTS}{CTS}{Clock Tree Synthesis}
\newacronym{DLP}{DLP}{Data Level Parallelism}
\newacronym{DMA}{DMA}{Direct Memory Access}
\newacronym{DMHLE}{DMHLE}{Dynamic Miss Handling Line Expansion}
\newacronym{DSE}{DSE}{Dynamic Subentry Expansion}
\newacronym{DRAM}{DRAM}{Dynamic Random-Access Memory}
\newacronym{DSA}{DSA}{Domain-Specific Accelerator}
\newacronym{DSP}{DSP}{Digital Signal Processing}
\newacronym{DUT}{DUT}{Device Under Test}
\newacronym{DOTP}{DotP}{Dot Product}
\newacronym{ECL}{ECL}{Emitter-Coupled Logic}
\newacronym{FBB}{FBB}{Forward Body-Biasing}
\newacronym{FC}{FC}{Fully-Connected}
\newacronym{FFT}{FFT}{Fast Fourier Transform}
\newacronym{FDSOI}{FD-SOI}{Fully Depleted Silicon on Insulator}
\newacronym{FMA}{FMA}{Fused Multiply-Add}
\newacronym{FPGA}{FPGA}{Field-Programmable Gate Array}
\newacronym{FPU}{FPU}{Floating Point Unit}
\newacronym{GEMM}{GEMM}{General Matrix Multiply}
\newacronym{GPGPU}{GPGPU}{General-Purpose \acrlong{GPU}}
\newacronym{GPU}{GPU}{Graphics Processing Unit}
\newacronym{HBM}{HBM}{High-Bandwidth Memory}
\newacronym{HDL}{HDL}{Hardware Description Language}
\newacronym{HERO}{HERO}{Heterogeneous Embedded Research Platform}
\newacronym{HPC}{HPC}{High-Performance Computing}
\newacronym{IoT}{IoT}{Internet of Things}
\newacronym{ILP}{ILP}{Instruction Level Parallelism}
\newacronym{IOT}{IoT}{Internet-of-Things}
\newacronym{IPC}{IPC}{Instructions Per Cycle}
\newacronym{IPU}{IPU}{Image Processing Unit}
\newacronym{ISA}{ISA}{Instruction Set Architecture}
\newacronym{LSU}{LSU}{Load/Store Unit}
\newacronym{LLC}{LLC}{Last Level Cache}
\newacronym{LLM}{LLM}{Large Language Model}
\newacronym{LVT}{LVT}{low voltage threshold}
\newacronym{MATMUL}{MatMul}{Matrix Multiplication}
\newacronym{GE}{GE}{Gate Equivalents}
\newacronym{MHA}{MHA}{Miss Handling Architecture}
\newacronym{MIMD}{MIMD}{multiple instruction, multiple data}
\newacronym{ML}{ML}{Machine Learning}
\newacronym{MMA}{MMA}{Matrix-Multiply Assist}
\newacronym{MME}{MME}{Matrix Multiplication Extension}
\newacronym{MMU}{MMU}{Memory Management Unit}
\newacronym{MRF}{MRF}{Matrix Register File}
\newacronym{MSHR}{MSHR}{Miss Status Handling Register}
\newacronym{MUL}{MUL}{multiplier}
\newacronym{MVL}{MVL}{maximum vector length}
\newacronym{NUMA}{NUMA}{Non-Uniform Memory Access}
\newacronym{NOC}{NoC}{Network-on-Chip}
\newacronym{NI}{NI}{Network Interface}
\newacronym{SEE}{SEE}{Single-Event Effect}
\newacronym{SEU}{SEU}{Single-Event Upset}
\newacronym{SET}{SET}{Single-Event Transient}
\newacronym{FF}{FF}{Flip-Flop}
\newacronym{MX}{MX}{Matrix eXtension}
\newacronym{PCIe}{PCIe}{Peripheral Component Interconnect Express}
\newacronym{PC}{PC}{Program Counter}
\newacronym{PE}{PE}{processing element}
\newacronym{PiM}{PiM}{Processing in memory}
\newacronym{PL}{PL}{Programmable Logic}
\newacronym{PMCA}{PMCA}{Programmable Manycore Accelerator}
\newacronym{PnM}{PnM}{Processing near memory}
\newacronym{PNR}{PnR}{Place-and-Route}
\newacronym{PSL}{PSL}{Power Service Layer}
\newacronym{PTE}{PTE}{page-table entry}
\newacronym{PTW}{PTW}{page-table walker}
\newacronym{PULP}{PULP}{Parallel Ultra Low Power}
\newacronym{RAW}{RAW}{read-after-write}
\newacronym{RBB}{RBB}{Reverse Body-Biasing}
\newacronym{ROB}{ROB}{Reorder Buffer}
\newacronym{RTL}{RTL}{Register Transfer Level}
\newacronym{RVT}{RVT}{Regular Voltage Threshold}
\newacronym{RVV}{RVV}{RISC-V Vector}
\newacronym{RoCC}{RoCC}{Rocket Custom Coprocessor Interface}
\newacronym{SCM}{SCM}{Storage Class Memory}
\newacronym{SIMD}{SIMD}{single instruction, multiple data}
\newacronym{SIMT}{SIMT}{single instruction, multiple thread}
\newacronym{SLDU}{SLDU}{Slide Unit}
\newacronym{LRU}{LRU}{Least Recently Used}
\newacronym{DS-LRU}{DS-LRU}{Dynamic Subset Least Recently Used}
\newacronym{SLVT}{SLVT}{super-low voltage threshold}
\newacronym{SM}{SM}{Streaming Multiprocessor}
\newacronym{SME}{SME}{Scalable Matrix Extension}
\newacronym{SOC}{SoC}{System-on-Chip}
\newacronym{SRAM}{SRAM}{Static Random-Access Memory}
\newacronym{SSE}{SSE}{Streaming SIMD Extension}
\newacronym{SVE}{SVE}{Scalable Vector Extension}
\newacronym{TCDM}{TCDM}{Tightly Coupled Data Memory}
\newacronym{TLP}{TLP}{Thread Level Parallelism}
\newacronym{TxnID}{TxnID}{Transaction ID}
\newacronym{VAC}{VAC}{Vector Access}
\newacronym{VC}{VC}{virtual channel}
\newacronym{VCONV}{VCONV}{Vector Conversion}
\newacronym{VEX}{VEX}{Vector Execute}
\newacronym{VFU}{VFU}{vector functional unit}
\newacronym{VID}{VID}{Vector Instruction Decode}
\newacronym{VIS}{VISSUE}{Vector Instruction Issue}
\newacronym{VLEN}{VLEN}{vector length}
\newacronym{VLIW}{VLIW}{Very Long Instruction Word}
\newacronym{VLOOP}{VLOOP}{Vector Loop}
\newacronym{VLR}{VLR}{vector length register}
\newacronym{VLSU}{VLSU}{Vector Load/Store Unit}
\newacronym{VNB}{VNB}{Von Neumann Bottleneck}
\newacronym{VRF}{VRF}{Vector Register File}
\newacronym{VPU}{VPU}{Vector Processing Unit}
\newacronym{VT}{VT}{vector thread}
\newacronym{WAR}{WAR}{write-after-read}
\newacronym{WAW}{WAW}{write-after-write}
\newacronym{DCT}{DCT}{discrete cosine transform}
\newacronym{TSV}{TSV}{through-silicon via}
\newacronym{3DIC}{3D-IC}{three-dimensional integrated circuit}
\newacronym{PPA}{PPA}{power, performance, and area}
\newacronym{F2F}{F2F}{face-to-face}
\newacronym{W2W}{W2W}{wafer-to-wafer}
\newacronym{IC}{IC}{integrated circuit}
\newacronym{C4}{C4}{controlled collapse chip connection}
\newacronym{FEOL}{FEOL}{front end of the line}
\newacronym{BEOL}{BEOL}{back end of the line}
\newacronym{PDP}{PDP}{power-delay product}
\newacronym{EDP}{EDP}{energy-delay product}
\newacronym{DRV}{DRV}{design rule violation}
\newacronym{DDR}{DDR}{double data rate}
\newacronym{SDRAM}{SDRAM}{synchronous dynamic random-access memory}
\newacronym{TPU}{TPU}{Tensor-Processing Unit}
\newacronym{FSM}{FSM}{Finite-State Machine}
\newacronym{VCD}{VCD}{Value Change Dump}
\newacronym{FIFO}{FIFO}{First-In First-Out}
\newacronym{PnR}{PnR}{Place-and-Route}
\newacronym{MoE}{MoE}{Mixture-of-Expert}
\def\BibTeX{{\rm B\kern-.05em{\sc i\kern-.025em b}\kern-.08em
    T\kern-.1667em\lower.7ex\hbox{E}\kern-.125emX}}
\renewcommand{\baselinestretch}{0.988}
\begin{document}
\newif\ifreviewmode
\reviewmodefalse

\title{
Assessing Triple Modular Redundancy for  Wide-Link, Low-Latency NoC Routers: Reliability and Physical Design Challenges
\ifreviewmode
\else
\fi
}

\ifreviewmode
\author{\emph{Hidden for double-blind review purposes.}}
\else
\author{\IEEEauthorblockN{Chen Wu\,\orcidlink{0009-0006-5417-2870}\textsuperscript{1}, Michael Rogenmoser\,\orcidlink{0000-0003-4622-4862}\textsuperscript{1}, Luca Benini\,\orcidlink{0000-0001-8068-3806}\textsuperscript{1,2} and Angelo Garofalo\,\orcidlink{0000-0002-7495-6895}\textsuperscript{1,2}}
\IEEEauthorblockA{\textsuperscript{1}\textit{ETH Zürich, Zürich, Switzerland}, \textsuperscript{2}\textit{Università di Bologna, Bologna, Italy}}

\{chenwu,michaero,lbenini,agarofalo\}@iis.ee.ethz.ch
}
\fi




\maketitle
\AddToShipoutPictureBG*{%
  \AtPageLowerLeft{%
    \put(\LenToUnit{\dimexpr(\paperwidth-\textwidth)/2\relax},\LenToUnit{0.35in}){%
      \parbox{\textwidth}{\footnotesize\copyright~2026 IEEE. Personal use of this material is permitted. Permission from IEEE must be obtained for all other uses, in any current or future media, including reprinting/republishing this material for advertising or promotional purposes, creating new collective works, for resale or redistribution to servers or lists, or reuse of any copyrighted component of this work in other works.}%
    }%
  }%
}

\begin{abstract}
Protecting the Network-on-Chip (NoC) of physical-AI tile-based accelerators deployed in harsh environments against single-event effects (SEEs) is paramount for preventing NoC failures that can lead to deadlocks and silent data corruption (SDC). Prior work on reliable NoCs has largely focused on narrow links (e.g., 32-bit), deeply pipelined routers, and single-event upsets (SEUs). However, the state of the art has evolved toward low-latency NoC routers with ultra-wide links, implemented on advanced technology nodes and operating at frequencies above 1 GHz. We evaluate the cost and reliability trade-offs of implementing Triple Modular Redundancy (TMR) at three granularities (coarse, state-only, and full) for a 2-cycle-latency NoC router with 512-bit wide links. 
We carry out RTL-to-GDSII physical design in TSMC 7nm technology, as well as both RTL- and netlist-level SEU and SET fault injection campaigns. We evaluate the three TMR approaches in terms of reliability, cost, and physical design strategies, further extending the assessment from a standalone router to a full AI acceleration tile. Our results show that state-only and coarse-grained TMR do not provide sufficient protection against SEEs, whereas full TMR eliminates all observed failures across more than one million injected faults per experiment. Although the standalone full-TMR router incurs a 7.04x area overhead, this cost is drastically amortized once integrated into a complete AI accelerator tile with processors and local L1 memories: the same design adds only 16.8\% area and 15.2\% power consumption under a GEMM benchmark at the system level, with the critical path of the tile entirely unaffected. These results demonstrate that advanced technology nodes provide sufficient routing capacity to make full TMR a practical and deployable solution for protecting NoCs in Physical AI systems operating in harsh environments.

\end{abstract}
\glsresetall
\glsunset{AI}
\glsunset{RTL}
\glsunset{GEMM}
\begin{IEEEkeywords}
Fault tolerance,
Single-event upsets,
Network-on-chip
\end{IEEEkeywords}

\section{Introduction}

Tile-based many-\gls{PE} accelerators have become the dominant architectural template for scalable Physical AI systems~\cite{lie2024waferscale, vasiljevic2024blackhole, lee2024tensorcontraction}. In these architectures, each tile integrates compute processors and local memory with a \gls{NOC} router and a \gls{NI}. Tiles are usually arranged in a 2D mesh floorplan, so that the system can be scaled up by simple tile abutment and replication. The bandwidth and heterogeneous traffic patterns of \gls{AI} workloads, however, exceed what traditional narrow-link, multi-stage pipelined \gls{NOC} routers can sustain. Therefore, modern \glspl{NOC} are evolving toward low-latency routers with ultra-wide physical links~\cite{fischer2025floonoc}, an evolution enabled by the abundant routing resources offered by advanced technology nodes.

Tile-based Physical AI \glspl{SOC} are increasingly adopted in autonomous safety-critical systems operating in radiation-heavy environments such as space~\cite{iturbe2015use},
where radiation-induced \glspl{SEU} and \glspl{SET} challenge the continuous operational correctness these systems inherently demand. The \gls{NOC}, as their backbone interconnect, is therefore critical to overall reliability: a single undetected or uncorrected fault in the \gls{NOC} can corrupt data, deadlock communication, or hang the entire system.

Prior work addresses \glspl{NOC} reliability through fault-tolerant routing algorithms~\cite{feng2012addressing} that reroute around the faulty links (or routers) by exploiting path redundancy, or by relying on microarchitectural modifications, including resource sharing~\cite{fick2009vicis, khalil2024dynamic}, bypass paths~\cite{fick2009vicis}, reconfigurable link designs~\cite{xu2023rmc_noc}, and retransmission paired with lightweight detection~\cite{constantinides2006bulletproof, feng2012addressing}. 
These works mainly target permanent faults from aging or manufacturing defects, and show one or more of three limitations: 1) they incur system-level performance drop due to increased latency and reduced throughput from traffic hotspots, which is especially detrimental for Physical AI systems where real-time constraints are tight, and frequent and bulk data movements across tiles can easily turn the \gls{NOC} into the dominant performance bottleneck once its latency degrades; 2) they depend on offline detection such as built-in self-test (BIST) and therefore cannot react timely to transient faults; 3) they introduce architecture-specific solutions (e.g., custom link designs) that bind the solution to one particular router microarchitecture and hinder broader adoption. Hybrid designs pairing error-correcting codes (ECC) with TMR, targeting spacecraft computing platforms~\cite{rausch2025reliability}, have also been explored, but they cover only \glspl{SEU} and apply TMR exclusively to FSM state and ECC to payload data, leaving a protection gap for control-adjacent sequential elements such as buffer pointers.
%
%
%
Moreover, almost all of these works target narrow-link, multi-stage pipelined routers, lack a thorough physical design evaluation, and evaluate the reliability through either analytical modeling or fault injection campaigns too limited to reach statistically meaningful fault coverage.

TMR, a well-established technique for real-time correction of transient faults~\cite{walsemann2023radiation, andorno2023radiation}, is well-suited to close these gaps: it corrects faults online; it is architecture-agnostic; and depending on the granularity, it can protect sequential and combinational logic against \glspl{SEE}. However, its cost–reliability trade-off has never been characterized on wide-link, low-latency \gls{NOC} routers,
where wide datapaths and short pipelines amplify the \gls{PPA} impact of TMR.
In this work, we characterize this trade-off by applying TMR at three granularities to FlooNoC~\cite{fischer2025floonoc}, a state-of-the-art open-source wide-link, low-latency \gls{NOC}, and evaluate the resulting designs in reliability, physical-design feasibility, and \gls{PPA} cost. To our knowledge, this is the first \gls{RTL}-to-GDSII characterization of TMR on such routers in an advanced technology node, backed by large-scale fault injection experiments. The contributions of this work are as follows:
 
\begin{itemize}
\item At the \gls{RTL} level, we implement three TMR schemes at different granularities on the FlooNoC router~\cite{fischer2025floonoc}, a 2-cycle-per-hop \gls{NOC} with parallel 512-bit wide and 64-bit narrow physical links.
\item We evaluate the performance-power-area costs and physical design challenges by implementing the three TMR routers and the baseline in TSMC 7nm technology, down to GDSII layout. We show that physical costs scale sharply with the granularity of redundancy. In particular, the per-flip-flop voter triplets of full TMR introduce dense local wiring that induces severe routing congestion, forcing the backend to absorb the entire timing closure burden during routing and pushing the maximum frequency down to nearly half of the baseline with a 7.04$\times$ area overhead.
\item We evaluate the reliability of the three TMR routers by performing exhaustive fault-injection campaigns, covering \glspl{SEU} and \glspl{SET}, isolated and multiple faults. We show that full TMR is the only solution to achieve full resilience to the \gls{SEE}-intensive radiation environment, while coarse TMR fails to recover from multiple accumulated faults irrespective of their timing, and state TMR from \glspl{SET};
\item We evaluate the three TMR routers at the system level by integrating them into a Physical AI tile~\cite{zaruba2020snitch} implemented in TSMC 7nm. We show that the standalone costs drastically reduce at the system level, where even the most costly full TMR router amortizes to just 16.8\% additional area and 15.2\% additional power consumption over the baseline tile. Moreover, the critical path of the tile is unaffected. This demonstrates that TMR is an expensive but viable solution to protect highly critical components of a Physical AI system deployed in harsh environments, like the \gls{NOC}.

\end{itemize}

\section{Background}

\subsection{Fault Model}
\label{sec:background:fault_model}
In this work, we focus on transient single-event faults (\glspl{SEU} and \glspl{SET}) and their accumulation over time.

A \gls{SEU} is a bit-flip in a state-saving element, primarily \glspl{FF} and SRAM cells, induced by a single ionizing particle strike. The corrupted bit persists until the storage element is overwritten, and can propagate to the design's outputs as an observable error.
If left uncorrected, a single \gls{SEU} increases the likelihood of subsequent failures when additional \glspl{SEU} accumulate in the same design, which is known as an effect of fault accumulation.
Furthermore, as reported in~\cite{elash2025design}, multi-bit \glspl{SEU}, in which a single particle strike flips several neighboring \glspl{FF} within the same clock cycle, are becoming increasingly relevant in modern technology nodes.

A \gls{SET}, in contrast, is a transient voltage pulse induced in combinational logic by the charge collected from a single ionizing particle strike. While individual \glspl{SET} are short-lived glitches, the higher operating frequencies of modern designs in scaled technologies make such transients more likely to be latched into a downstream \gls{FF}, thereby manifesting as observable errors~\cite{dodd2003basic}.

\begin{figure}[!t]
    \centering
    \includegraphics[width=\linewidth, trim=0 6pt 0 6pt, clip]{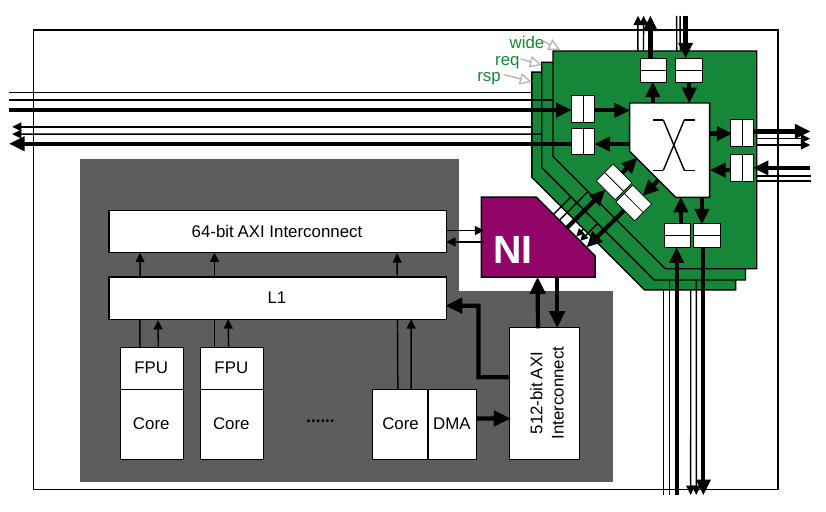}%
    \vspace{-3pt}%
    \caption{Architecture of a compute tile. FlooNoC components are marked as green (router) and purple (\gls{NI}).}
    \label{fig:bg}
\end{figure}

\subsection{FlooNoC} \label{sec:background:floonoc}

This work builds on FlooNoC~\cite{fischer2025floonoc},
a state-of-the-art, \gls{AXI}-compliant, open-source \gls{NOC}.
Unlike traditional multi-stage pipelined NoC routers, FlooNoC features only 2-cycle latency per hop, with no pipeline stages between the input and output buffers within the router. To efficiently sustain the heterogeneous traffic of modern \gls{AI} workloads, which mix wide burst data with latency-critical control messages, FlooNoC provides two parallel networks: a 512-bit \emph{wide} network for burst data and a 64-bit \emph{narrow} network for control messages.

As shown in \autoref{fig:bg}, the FlooNoC router has three physical links: one wide link for bursted read/write data transfers and two narrow links for latency-critical requests and responses. Each link has its own router, enabling fully decoupled traffic and avoiding frequent stalls and deadlocks. The multi-link router connects to endpoints through a \gls{NI} that maps the \gls{AXI} transactions of the narrow and wide channels onto the flits of the three physical links of the \gls{NOC}.

\subsection{Compute Cluster}
\label{sec:background:cluster}


As a representative tile of a Physical AI acceleration engine, we adopt the Snitch Cluster~\cite{zaruba2020snitch}, integrated with FlooNoC as shown in \autoref{fig:bg}. Each cluster comprises eight energy-efficient RISC-V cores, each paired with a 64-bit SIMD-capable \gls{FPU}, and a shared 128\,KB L1 scratchpad memory. A ninth core is coupled with a \gls{DMA} engine that orchestrates bulk data transfers between the cluster and the rest of the system. This tightly coupled, scratchpad-centric design sustains the high throughput required by modern \gls{AI} workloads with competitive energy efficiency. The cluster's internal 64-bit and 512-bit \gls{AXI} interconnects are routed to an \gls{NI}, which maps them onto the narrow request, narrow response, and wide data channels of the multi-link router. Together, the cluster, the \gls{NI}, and the multi-link router constitute a compute tile of a Physical AI accelerator system.

We use this Snitch-based cluster tile to evaluate the proposed TMR FlooNoC routers at the system level.
\section{Implementation} \label{sec:implementation}

\begin{figure}[t]%
\centering%
\subfloat[\label{fig:tmr_coarse}Coarse TMR.]{%
\includegraphics[height=0.28\linewidth]{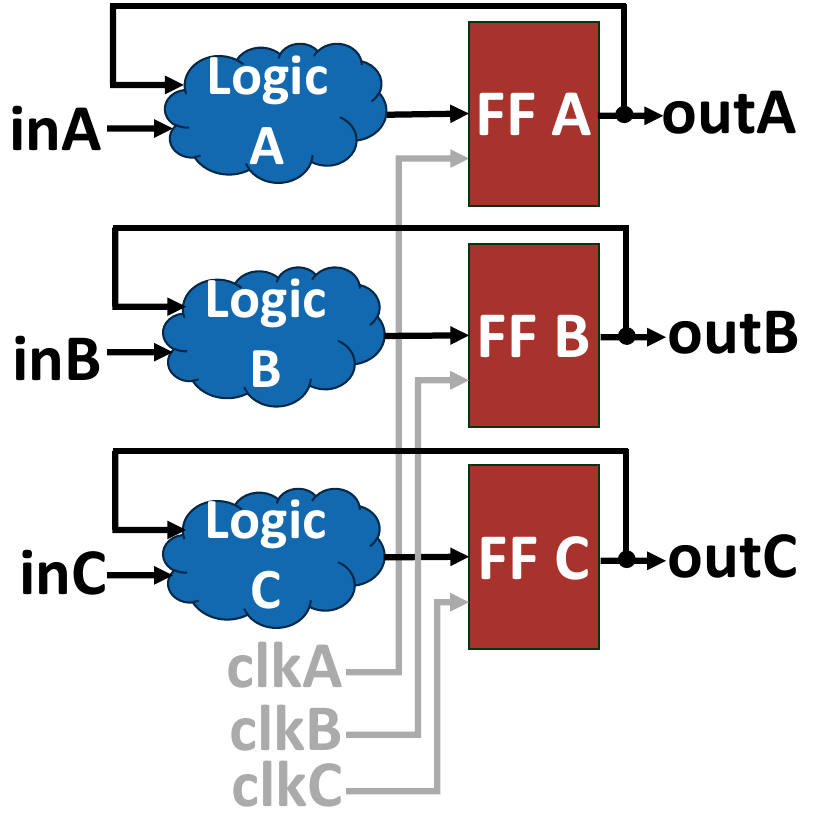}%
}%
\hfill%
\subfloat[\label{fig:tmr_state}State-only TMR.]{%
\includegraphics[height=0.27\linewidth]{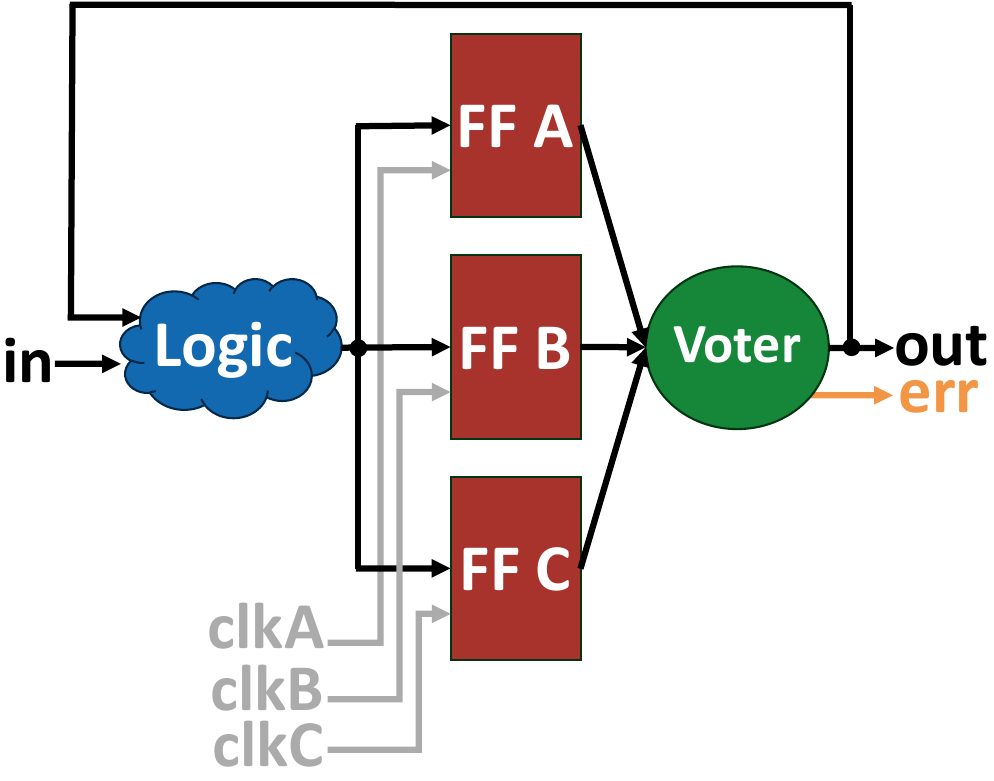}%
}%
\hfill%
\subfloat[\label{fig:tmr_full}Full TMR.]{%
\includegraphics[height=0.28\linewidth,]{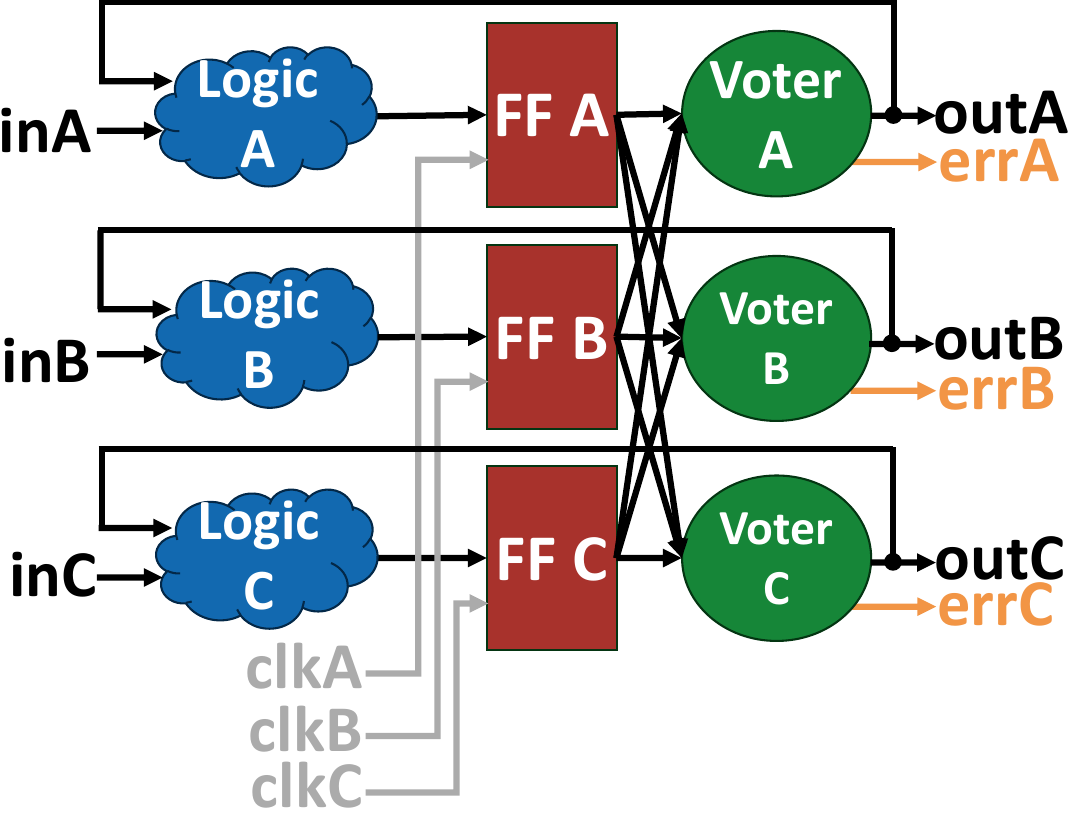}%
}%
\caption{Illustration of the three TMR schemes.}%
\label{fig:tmr}%
\end{figure}

We implement three TMR routers at different granularities. Because the router's internal logic is complex, we illustrate the three schemes on a generic diagram of basic combinational and sequential elements, as shown in \autoref{fig:tmr}.

\textbf{Coarse TMR}: The entire router is triplicated by instantiating three replicas of the baseline design. The three replicas are identical and operate in parallel, with no interaction between them. The flits are therefore triplicated and routed over independent physical links, with majority voting performed outside the router at the \gls{NI} of the destination endpoint. Compared to hop-by-hop voting, where a per-hop voter that consolidates the three replicas at every router would itself become a single point of failure on the data path, this end-to-end scheme keeps the three flit copies on fully isolated paths throughout the network.

\textbf{State-only TMR}: We triplicate only the sequential elements of the router, including control \glspl{FF} and data buffers, while the combinational logic (e.g., routing computation) remains shared across the triplicated sequential elements. As illustrated in Figure\autoref{fig:tmr_state}, each triplet of state elements drives a single majority voter, whose voted output feeds the shared combinational fabric. 
Each internal voter also raises an error signal when its three inputs disagree. We OR all such signals together and expose the result at the router's output as a single-bit aggregated error signal, which a system-level reliability monitor can use to track the per-router \gls{SEU} rate.

State-only TMR recovers any \gls{SEU} that would otherwise corrupt the latched state, propagate downstream, and potentially manifest as a system-level error, but does not protect \glspl{SET} in the combinational logic, trading \gls{SET} coverage for lower hardware complexity and overhead. This trade-off is justified by the observation that most combinational \glspl{SET} dissipate before the next clock edge and are never captured by a flip-flop~\cite{buchner1997comparison}.

\textbf{Full TMR}: In addition to the triplication and voting mechanism adopted in state-only TMR, we also triplicate the combinational logic of the router. As illustrated in Figure\autoref{fig:tmr_full}, each \gls{FF} thus has its own dedicated combinational logic and voter, so that a fault in the combinational logic of one replica cannot affect the other two or compromise the correction mechanism. Because this method propagates triplicated flits at the router output, full TMR also requires voters at the \gls{NI}, as in coarse TMR. Similar to the state-only TMR, the aggregated error signal is also exposed at the router's output. Together, the per-\gls{FF} internal voters, the triplicated combinational logic, and the end-to-end \gls{NI} voting ensure that any single fault, whether in a sequential or combinational element, is confined to a single replica and corrected on the next clock edge, leaving no opportunity for cross-cycle accumulation. This makes full TMR the most reliable of the three variants.

\begin{figure}[t]%
\centering%
\subfloat[\label{fig:cg_base}Clock-gated \gls{FF} in baseline router.]{%
\begin{minipage}[b]{0.34\linewidth}\raggedleft
\includegraphics[height=55.4pt]{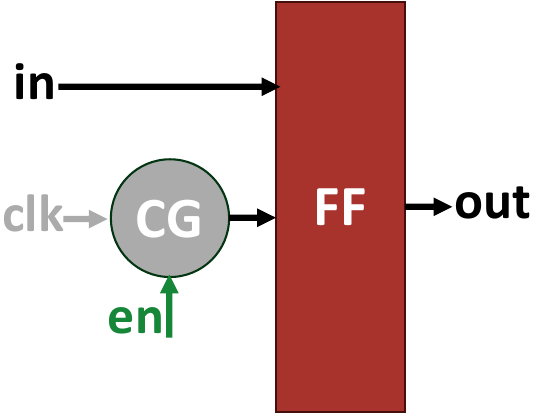}
\end{minipage}}%
\hfill%
\subfloat[\label{fig:cg_tmr}Clock-gated \gls{FF} with feedback path from the voter in TMR router.]{%
\begin{minipage}[b]{0.62\linewidth}\centering
\includegraphics[height=60.5pt, trim=16pt 0 15pt 14pt, clip]{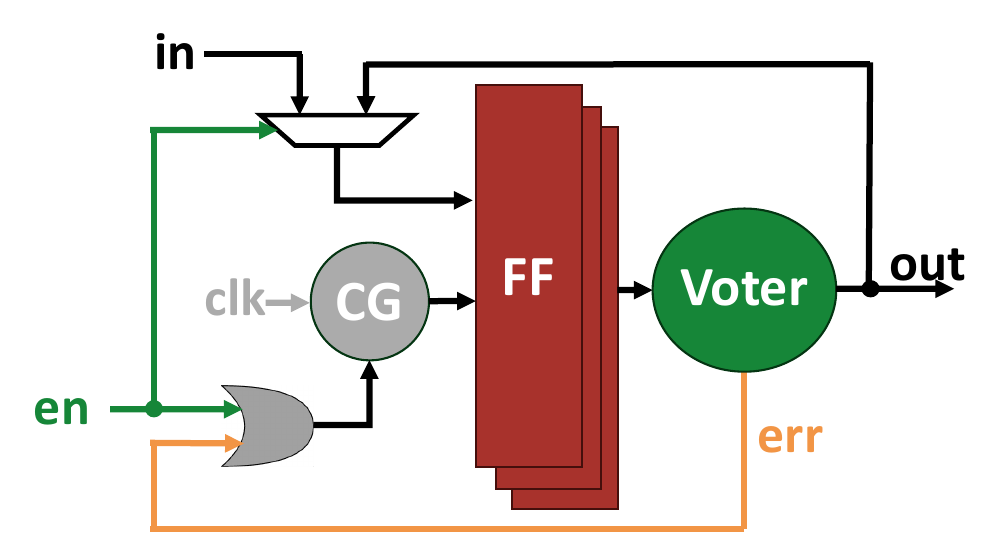}
\end{minipage}}%
\caption{Illustration of clock gating scheme for TMR routers.}%
\label{fig:cg}%
\end{figure}

Some \glspl{FF} in the baseline router are clock-gated to save power, as shown in Figure\autoref{fig:cg_base}. For state-only and full TMR routers, which contain internal voters, we feed each internal voter's error signal back into the clock-gating logic so that a corrupted \gls{FF} can still be promptly corrected even when its clock is gated, as illustrated in Figure\autoref{fig:cg_tmr}.

Since both state-only and full TMR require fine-grained triplication of the internal router components, we use the open-source TMRG tool~\cite{kulis2017single} to automatically insert TMR logic into the baseline router at the \gls{RTL} level. Moreover, all three TMR routers drive the three \gls{FF} replicas from independent clock and reset sources, kept separate down to the router boundary, preventing a common-mode glitch on either signal from defeating the voting and correction mechanism.

Both coarse and full TMR additionally require border voters at the \gls{NI} of the destination endpoint to perform majority voting on the triplicated flits. We focus on the router, the most critical \gls{NOC} component, and leave the \gls{NI}, where these voters reside, outside our evaluation: applying the same TMR methodology to the \gls{NI}, which is only about one sixth of the router area, is a straightforward extension from the methodological viewpoint.
\section{Physical Design of TMR Routers}

To quantify the area and timing costs of the three TMR variants and to assess the physical design challenges they introduce, we carry out the full physical implementation flow for the baseline, coarse-TMR, state-only-TMR, and full-TMR routers in \textsc{TSMC} \SI{7}{\nano\meter} technology with \textsc{Fusion Compiler 2024.09}. All four designs are placed and routed in a die area sized to reflect their integration into a realistic compute tile, targeting worst-case conditions (SS, \SI{125}{\celsius}, \SI{0.675}{\volt}).


To preserve the TMR redundancy, we apply physical design constraints to avoid that: 1) replicas of each register are optimized into merged multi-bit registers with the same clock and reset signals; 2) boundary optimizations filter out the voters. 
Moreover, to prevent common-mode events such as multi-bit upsets, we apply a placement constraint enforcing a minimum Manhattan distance between the replicas of each register.





\begin{figure}[t]
    \centering
    \includegraphics[width=\columnwidth, trim=0 6pt 0 14pt, clip]{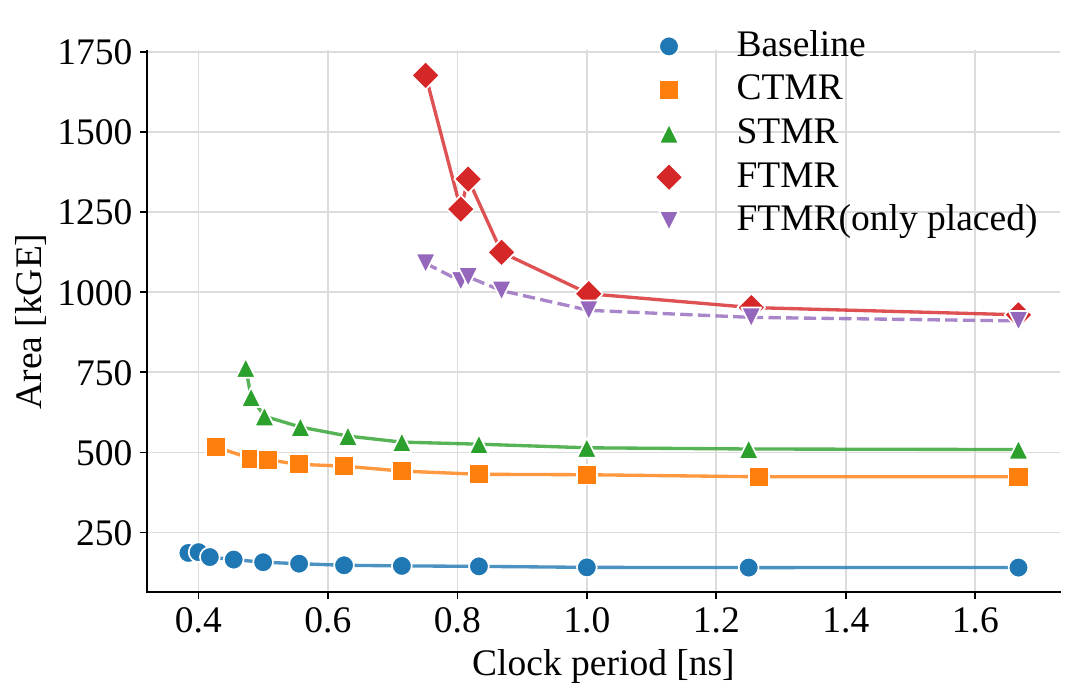}%
    \vspace{-3pt}%
    \caption{Area-Timing plot of different router implementations.}
    \label{fig:at}
\end{figure}

For each design, we sweep the target frequency, and we report the area-timing trade-off in \autoref{fig:at}.

In terms of timing, the coarse TMR router achieves a maximum frequency of \SI{2.34}{\giga\hertz}, with only a \SI{10}{\percent} degradation compared to the baseline's \SI{2.6}{\giga\hertz}. 
The state-only TMR router degrades further to \SI{2.11}{\giga\hertz} due to the additional internal voter logic and the feedback path of the error signal back to the clock gate. As intuitively inspected, the full TMR router shows the most pronounced degradation, reaching a maximum of \SI{1.33}{\giga\hertz}. This degradation arises because the per-\gls{FF} voter triplets introduce dense local wiring that produces severe routing congestion. The purple dashed line in \autoref{fig:at} indicates that the area grows sharply during the routing stage, showing that placement fails to anticipate this routing difficulty and leaves the entire timing pressure to be resolved during routing. 

Comparing all designs at the \SI{1}{\giga\hertz} target frequency of the Physical AI tile we consider later in this work, full TMR incurs by far the largest area overhead: coarse TMR, state-only TMR, and full TMR occupy $3.05\times$, $3.64\times$, and $7.04\times$ the baseline area, respectively. Three factors account for the steep cost of full TMR. First, we not only triplicate all internal logic but also insert a majority voter after every \gls{FF}, and \glspl{FF} already dominate the input and output buffers that account for more than 70\% of the router area. Second, a large OR-tree is synthesized to aggregate the per-voter error signals and expose them at the output port. Third, and most importantly, the routing stage must absorb the entire timing optimization burden and therefore inserts many additional buffers to meet the timing target.

Taken together, these results show that the physical cost of TMR grows sharply with the granularity of redundancy. Whether such standalone overheads are warranted ultimately depends on the reliability gain they deliver and on how they amortize once the router is integrated into a complete compute tile, both of which we evaluate in the following sections.

\section{Reliability Evaluation}

\subsection{Fault Injection Setup} \label{sec:fi_methodology}
To evaluate the reliability of the three TMR variants implemented in the FlooNoC multi-link router, we conduct extensive fault injection campaigns with Synopsys\textregistered{} VC Z01X\texttrademark{}-2025.06, a commercial concurrent fault simulator that supports fault injection at both the \gls{RTL} and netlist levels.


\textbf{Testbench Architecture}: To fully exercise all input and output ports in all directions and across all three physical channels of the multi-link router, we build a custom testbench around a $3\times3$ mesh of endpoints.
Only the central router is the \gls{DUT} we consider for our fault injection campaigns.
We simulate 400 random transactions, uniformly distributed among narrow/wide read/write, that traverse the router under test.

\textbf{Fault Sampling}: VC Z01X\texttrademark{} samples faults from a fault universe defined by all possible injection sites within the \gls{DUT} and all possible time points throughout the testbench simulation. Since the candidate designs differ substantially in the cardinality of their fault universes, we apply statistical fault sampling at a 99\% confidence level with a 0.1\% confidence interval, yielding sufficient fault coverage for each router design and averaging over 1 million simulated faults per experiment.

\textbf{Outcome Classification}: Each injected fault is classified into one of three categories: \emph{Masked}, \emph{Corrected}, or \emph{Faulty}. 

Because the \gls{DUT} is an interconnect component, an outcome is labeled \emph{Faulty} only when the fault perturbs the transactions observable at the router boundary. Concretely, two classes of output flits qualify as \emph{Faulty}: flits with corrupted payloads or steered to an incorrect egress port, and missing flits that enter the router at an input port but never reach any output port by the end of the simulation. Conversely, transient deviations on auxiliary handshake signals that do not produce an erroneous handshake, such as a \texttt{ready} pulse asserted while a \texttt{valid} remains deasserted, are treated as benign interface glitches and classified as \emph{Masked}.

We early-terminate a fault simulation only after its outcome is classified as \emph{Faulty}; otherwise, the simulation continues until the testbench naturally completes. This prevents us from missing faulty cases in which a fault is temporarily masked, or appears corrected but eventually triggers an erroneous transaction elsewhere in the system.

\subsection{Single-Fault Origin Analysis}
The first set of experiments targets single-fault injections, in which each simulation injects exactly one fault into the \gls{DUT} at a random time point, to characterize the resilience of the three TMR variants to isolated \glspl{SEE}.

\begin{figure}[t]
    \centering
    \subfloat[\texttt{FLOP} faults on \gls{RTL}]{%
        \includegraphics[width=0.5\linewidth, trim=0 32pt 0 3pt, clip]{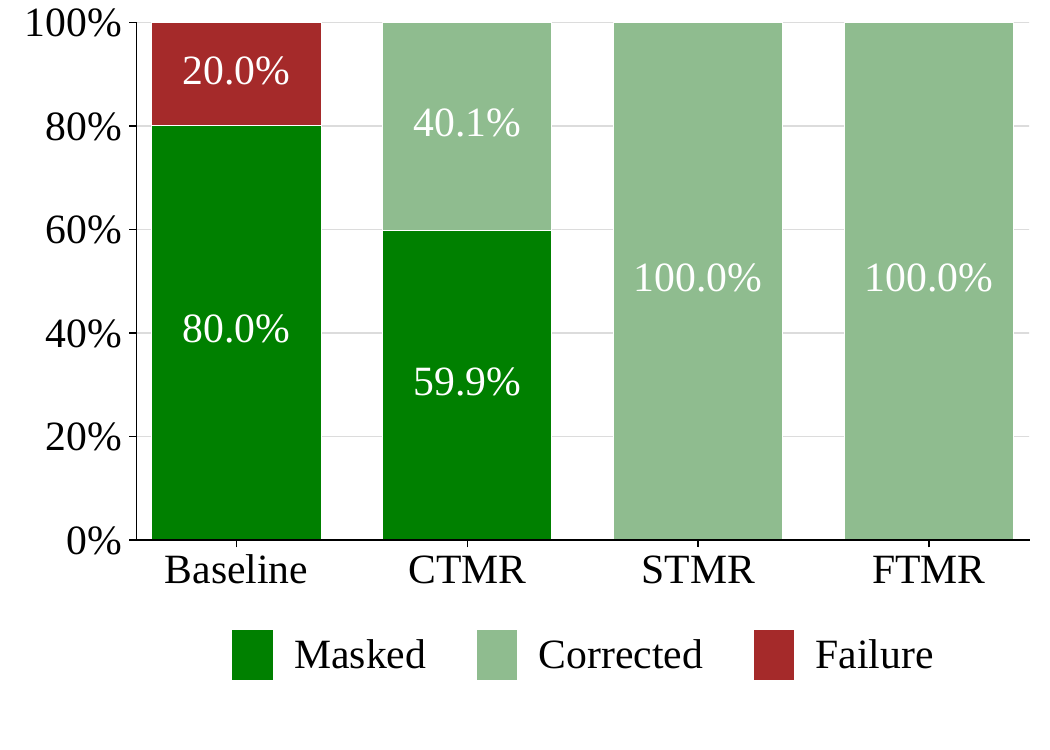}%
        \label{fig:sfo_flop}}
    \hfill
    \subfloat[\texttt{PRIM} faults on netlist]{%
        \includegraphics[width=0.5\linewidth, trim=0 2pt 0 3pt, clip]{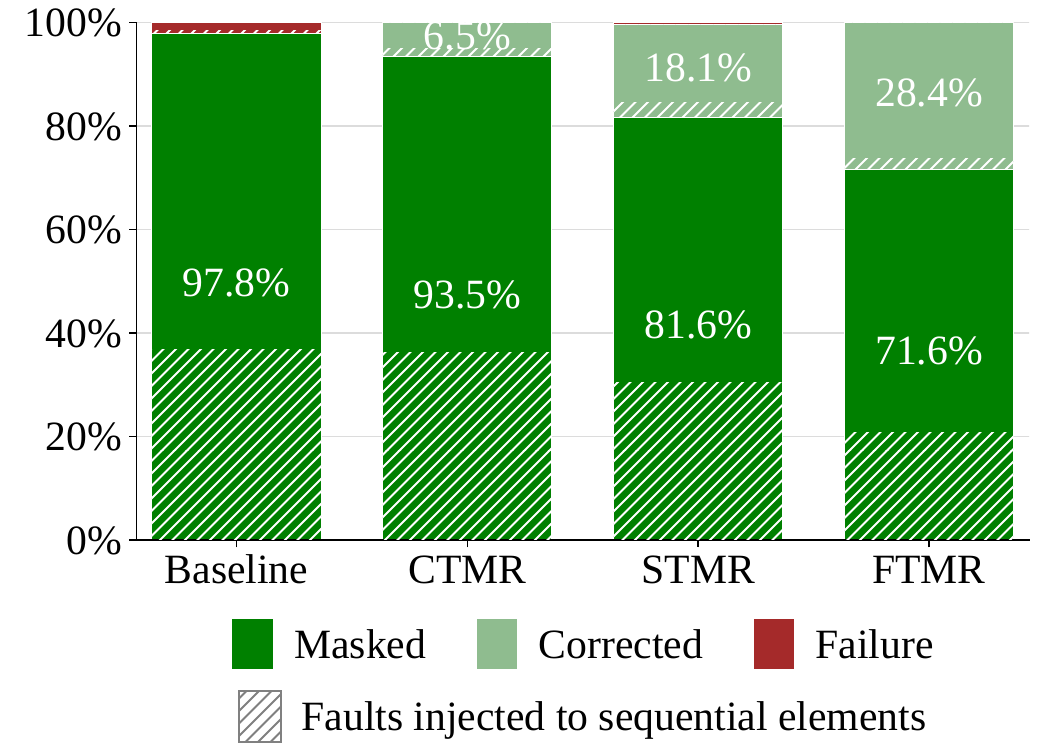}%
        \label{fig:sfo_prim}}
    \caption{Single-fault injection results.}
    \label{fig:sfo}
\end{figure}

We first emulate pure \glspl{SEU} by injecting \texttt{FLOP} (sequential-block) faults at the \gls{RTL} level, each flipping the value of a randomly selected register at a random cycle and holding it until a new value is latched.
Figure\autoref{fig:sfo_flop} shows that all three TMR variants exhibit a 0\% failure rate, in contrast to a whopping 20\% failure rate in the unprotected baseline design. 
Notably, a non-negligible fraction of faults in the coarse TMR router are classified as \emph{Masked}, unlike the 100\% \emph{Corrected} rate of the other two variants. This is because many faults do not propagate to the output ports and thus do not trigger the border voters.

To further evaluate resilience to transient \glspl{SEE}, we also inject \texttt{PRIM} (Verilog primitive) faults into the post-place-and-route netlists, flipping the output of a gate or a sequential element for a single cycle.
Figure\autoref{fig:sfo_prim} reports the resulting outcome distribution, where the white hatching on the bars denotes faults injected into sequential elements.
Faults injected into sequential elements are exclusively corrected or masked across all three TMR variants, consistent with the \texttt{FLOP} fault injection results.
Both coarse TMR and full TMR yield a 100\% combined \emph{Corrected}-or-\emph{Masked} outcome for every injected fault, although full TMR exhibits a higher proportion of \emph{Corrected} outcomes. This is because the internal voters of full TMR correct mismatches inside the design, even when those mismatches would have been inherently masked before reaching the strobed output ports. In contrast, the state-only TMR exhibits a residual 0.3\% failure rate: faults injected into the combinational logic can be latched and shared across all three flip-flop replicas, which constitutes an intrinsic vulnerability of the state-only TMR design.
Even this small residual rate rules out state-only TMR for harsh environments, where sustained particle flux makes such failures inevitable over the mission lifetime.

\subsection{Multi-Fault Origin Analysis}

We assess the resilience of the three TMR variants under multi-fault scenarios. As the size of the multi-fault universe grows exponentially with the number of concurrent faults, exhaustive sampling becomes expensive in terms of fault injection campaign runtime. We therefore restrict this campaign to two-fault simulations in which exactly two \texttt{FLOP} faults are injected per simulation, with the injection sites confined to critical control \glspl{FF}. Data buffers are excluded from the campaign because they are continuously refreshed by incoming flits and are therefore far less prone to fault accumulation, as confirmed by the analysis at the end of this subsection.

\begin{figure}[t]
    \centering
    \subfloat[Same clock cycle]{%
        \includegraphics[width=0.5\linewidth, trim=0 32pt 0 3pt, clip]{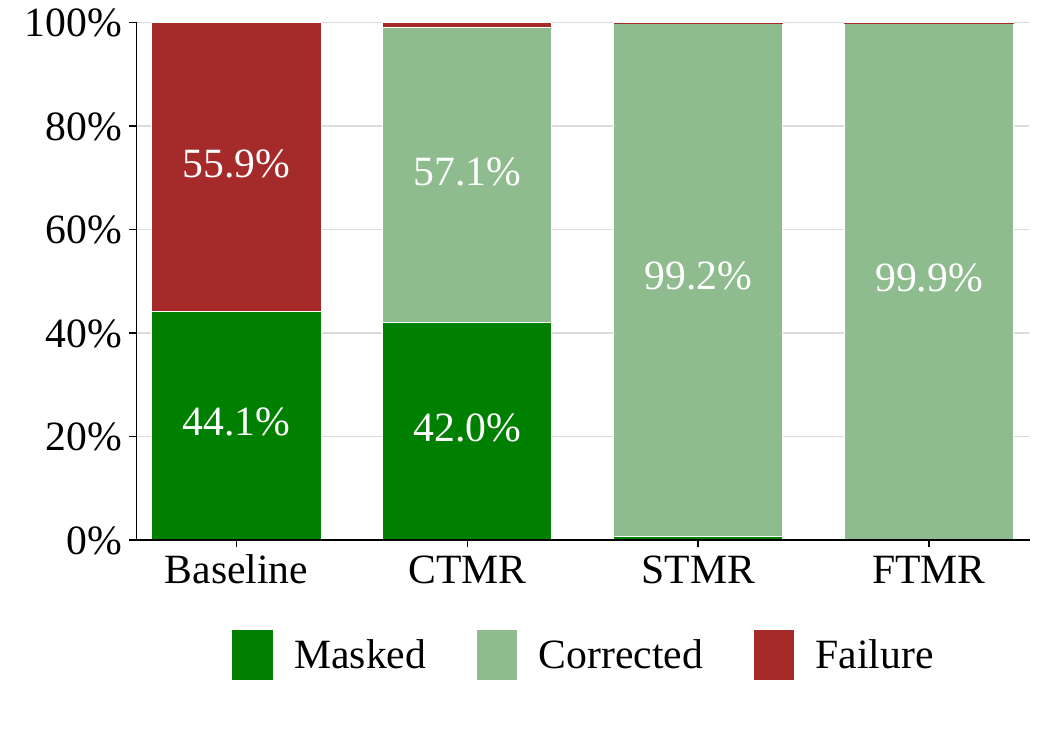}%
        \label{fig:mfo_samecycle}}
    \hfill
    \subfloat[Random cycles, random positions]{%
        \includegraphics[width=0.5\linewidth, trim=0 32pt 0 3pt, clip]{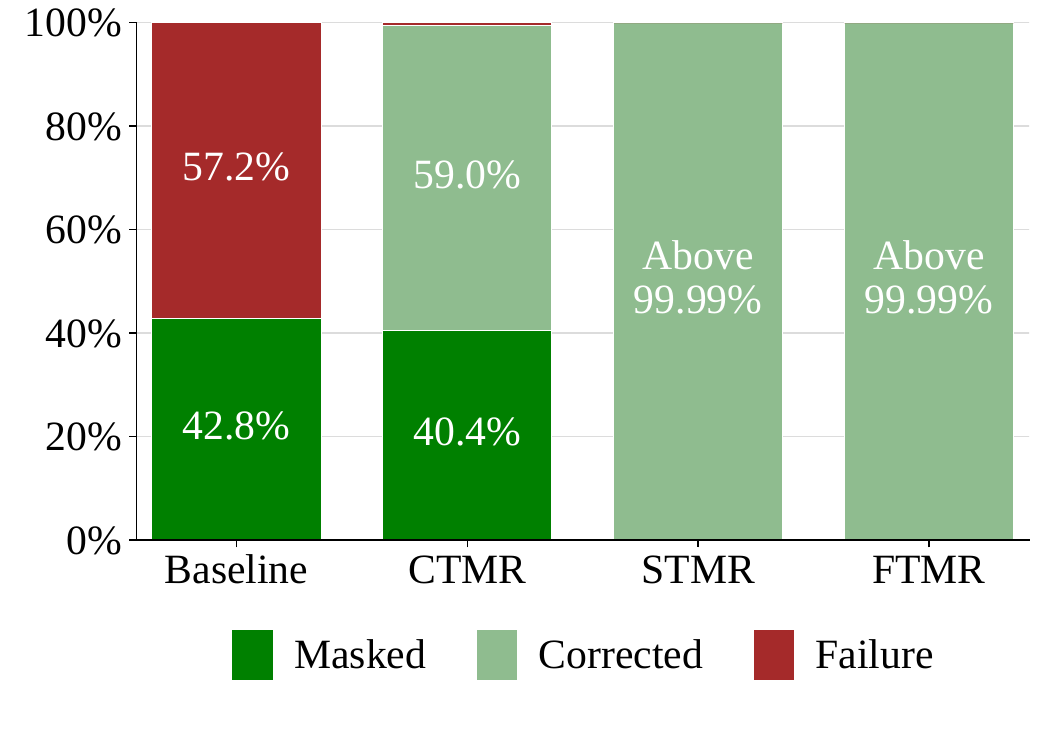}%
        \label{fig:mfo_rand_cycle_rand_pos}}
    \caption{Multi-fault injection results.}
    \label{fig:mfo}
\end{figure}

In the first experiment, the two faults are injected within the same clock cycle but into different \glspl{FF} (Figure\autoref{fig:mfo_samecycle}). The unprotected baseline exhibits a substantially higher failure rate than in the single-fault case. All three TMR variants produce a small number of \emph{Faulty} outcomes at comparable failure rates below 1\%, and every failure arises when the two faults coincide on two replicas of the same \gls{FF}. This is an expected fundamental limitation of the TMR methodology against multi-bit errors that simultaneously corrupt two of the three replicas.

In the second experiment, the two faults are injected at random cycles. As the fault universe size would become too large and fault simulations would become too slow if the two faults could occur at arbitrary time points across the whole testbench, we randomly selected a 1000-cycle window as the possible injection time period. Results in Figure\autoref{fig:mfo_rand_cycle_rand_pos} show that the state-only TMR and the full TMR yield fewer than ten failures out of more than 1.6 million injected faults, all attributable to the same rare event as in the first experiment: the two faults coincide on two replicas of the same \gls{FF} within the same cycle. The coarse TMR, although capable of correcting the first injected fault, exhibits a substantially higher failure rate than the other two TMR variants, with failures persisting even when the two faults occur in different cycles and on different \glspl{FF}. This reveals the severe vulnerability of the coarse TMR design to fault accumulation. Since coarse TMR provides no self-recovery mechanism, the first injected fault may persist within the design and drive the affected replica router to diverge from the other two; any subsequent fault that corrupts either of the remaining replicas, regardless of its location, therefore causes the entire TMR scheme to fail.

\begin{figure}[t]
    \centering
    \includegraphics[width=\linewidth, trim=0 18pt 0 4pt, clip]{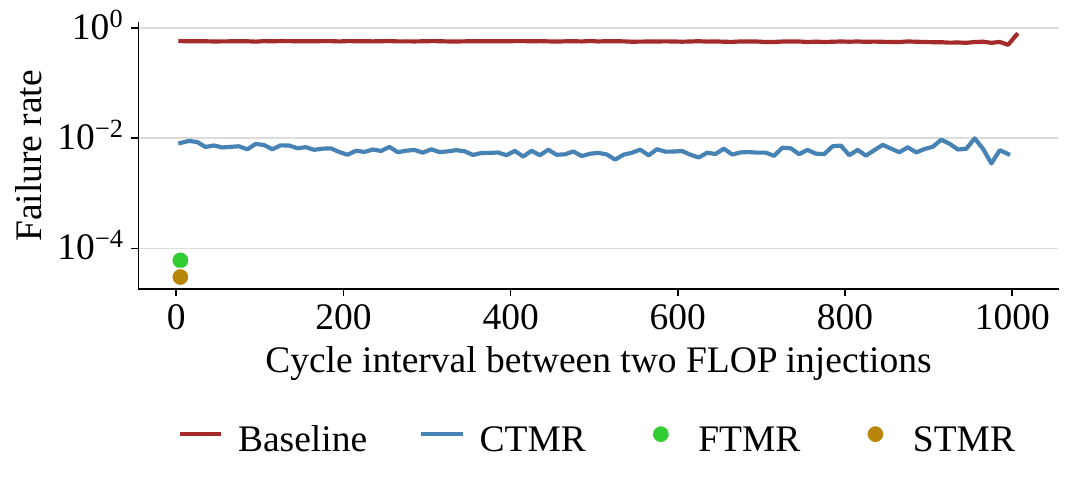}%
    \vspace{-3pt}%
    \caption{Failure rate versus the time interval between the two injected faults.}
    \label{fig:failurerate_interval}
\end{figure}

To further characterize this vulnerability of the coarse TMR design, we report the failure rate as a function of the time interval between the two injected faults in \autoref{fig:failurerate_interval}.
The failure rate of coarse TMR under accumulated multi-bit faults is essentially independent of the time interval. The reason is that critical control \glspl{FF}, unlike data buffers, are not necessarily resynchronized with the other two replicas by incoming flits after corruption. For instance, corrupting the \gls{FF} that stores the \emph{full} signal of an input buffer can stall the affected replica router indefinitely, since no incoming flit is accepted at that input port. Likewise, corrupting the \gls{FF} that stores the \emph{valid} signal of an output flit awaiting arbitration can temporarily stall the output port and cause the affected replica to diverge from the other two over the subsequent cycles. The failure rate of coarse TMR, therefore, depends solely on whether the two faults compromise two distinct replica routers, irrespective of when each fault occurs. In conclusion, these multi-fault experiments establish per-cycle internal correction as the decisive feature for sustained reliability under fault accumulation.

When projecting these per-fault outcome distributions to operational failure rates, the results in \autoref{fig:sfo} and \autoref{fig:mfo} should be weighted by each design's area, since a larger design intercepts proportionally more \glspl{SEE} under a given particle flux. Such weighting does not alter the comparison: full TMR corrects every single fault regardless of its larger cross-section, and its only residual failure mode, two faults striking two replicas of the same \gls{FF} in the same cycle, is largely suppressed by the enforced replica separation.
\section{System Evaluation}

\label{sec:results:system}

We further implement the full compute tile, introduced in \autoref{sec:background:cluster} but here integrated with configurable TMR routers, in \textsc{TSMC} \SI{7}{\nano\meter} technology with \textsc{Fusion Compiler 2024.09}, in order to study its system-level impact.

All three TMR variants meet the \SI{1}{\giga\hertz} frequency target under worst-case conditions (SS, \SI{125}{\celsius}, \SI{0.675}{\volt}), matching the baseline cluster with no observable timing degradation. This confirms that, although loss in maximum frequency when implemented standalone, this degradation does not manifest at the system level: the critical path lies inside the processor core rather than in the router, and the system target frequency lies well below the maximum frequency that even the slowest standalone TMR router can sustain.

\autoref{tab:sysresults} reports the area of each implementation along with its power consumption on a representative FP64 \gls{GEMM} workload, breaking down both metrics into the cluster-tile total and the router's contribution. The state-only TMR router consumes much less switching power because there is no triplication of combinational logic and module ports.

\begin{table}[t]
    \centering
    \caption{Area and power of the compute tile with different TMR routers.}
    \label{tab:sysresults}
    \setlength{\tabcolsep}{4.5pt}
    \begin{tabular}{lllll}
        \toprule
        \multirow{2}{*}{\textbf{Router}} & \multicolumn{2}{c}{\textbf{Area (kGE)}} & \multicolumn{2}{c}{\textbf{Power (mW)}} \\
        & \multicolumn{1}{c}{Total} & \multicolumn{1}{c}{Router} & \multicolumn{1}{c}{Total} & \multicolumn{1}{c}{Router} \\
        \midrule
        Baseline & 5006           & 161         & 83.8           & \phantom{0}3.2 \\
        CTMR &     5286 (+5.60\%) & 443 (2.75x) & 93.3 (+11.3\%) & 11.7 (3.69x) \\
        STMR &     5337 (+6.62\%) & 526 (3.26x) & 87.0 (+3.8\%) & \phantom{0}6.2 (1.96x) \\
        FTMR &     5847 (+16.8\%) & 965 (5.99x) & 96.5 (+15.2\%) & 11.6 (3.66x) \\
        \bottomrule
    \end{tabular}
\end{table}

Overall, the system-level \gls{PPA} overhead is affordable because the router occupies only a small fraction of the entire cluster tile.
Moreover, because the router itself is a single point of failure within the compute tile, spending a modest fraction of additional silicon to harden this is a worthwhile investment.
Consequently, even the full-TMR router, whose standalone overhead appears prohibitive, can be considered a viable choice at the system level on modern technology nodes for safety-critical applications and harsh environments.
Such amortization is specific to the compute tile and the \SI{1}{\giga\hertz} target considered here. The qualitative conclusion, however, holds broadly, as the router occupies only a small fraction of any realistic AI compute tile.



\section{Conclusions}
In this work, we conducted a thorough reliability and physical design analysis of three TMR approaches (coarse, state-only, and full) applied to wide-link, low-latency \gls{NOC} routers for Physical AI systems.
We characterized them through statistically meaningful \gls{RTL}- and netlist-level fault injection campaigns covering both \glspl{SEU} and \glspl{SET} under single- and multi-fault scenarios, complemented by a full \gls{RTL}-to-GDSII physical implementation in \textsc{TSMC} \SI{7}{\nano\meter} technology. 
At the router level, full TMR incurs the highest area and timing overhead but is the only scheme that achieves full resilience to the full spectrum of single-event effects expected in harsh environments, whereas coarse and state-only TMR, despite their lower cost, each leave residual vulnerabilities (fault accumulation across cycles and uncovered combinational \glspl{SET}, respectively). As inherent to any TMR design, the rare coincident upset striking two replicas of the same flip-flop remains uncorrectable, an event that the enforced spatial separation between replicas renders extremely unlikely.
When contextualized into a complete Physical AI compute tile, however, full TMR is amortized with a moderate area and power overhead with no impact on system performance. This makes router-level hardening a practical option for next-generation Physical AI systems deployed in harsh environments, ensuring reliable on-chip communication free of data corruption and deadlocks caused by \glspl{SEE}.

\bibliographystyle{IEEEtran}
\bibliography{main}



\end{document}